\documentclass[runningheads]{llncs}
\usepackage[T1]{fontenc}
\usepackage{graphicx}

\usepackage{amsthm}
\usepackage{graphicx}
\usepackage[many]{tcolorbox}
\usepackage{mathtools,amssymb,latexsym,amsfonts,stmaryrd}
\usepackage{pbox}
\usepackage{multirow}
\usepackage{tikz}
\usepackage{mathrsfs}
\usepackage{mathpartir}
\usepackage[ruled,linesnumbered]{algorithm2e}
\usepackage{url}
\usepackage{syntax}
\usepackage{framed}
\usepackage[noend]{algpseudocode}
\usepackage{flushend}
\usepackage{listings}
\usepackage{moresize}
\usepackage{wrapfig}
\usepackage{seqsplit}
\usepackage{alltt}
\usepackage{xspace}
\usepackage{hyperref}
\usepackage{subfigure}
\usepackage{enumitem}
\usepackage{pifont}

\DeclareMathAlphabet{\mathcal}{OMS}{cmsy}{m}{n}

\usepackage{amsmath, listings, amsthm, amssymb, proof, xspace}

\usepackage{thmtools}

\declaretheoremstyle[spaceabove=\topsep,notefont=\normalfont\itshape]{mystyle}

\newcommand{\revise}[2]{{\color{red}{\ifx&#1&\else- #1\fi}} {\color{ForestGreen}{\ifx&#2&\else+ #2\fi}}}%
\renewcommand{\revise}[2]{#2}%

\usetikzlibrary{arrows,patterns, decorations.pathreplacing}

\newcommand{\F}{Fig.}
\newcommand{\T}{Table}
\renewcommand{\S}{Sec.}

\newcommand{\ignore}[1]{}

\lstdefinestyle{base}{
  moredelim=**[is][\color{red}]{@}{@},
  escapeinside={<@}{@>}
}

\usepackage{amssymb}
\usepackage{pifont}

\newcommand\DejaVuttfamily{%
  \fontfamily{DejaVuSansMono-TLF}\selectfont
}

\lstdefinestyle{base}{
  moredelim=**[is][\color{red}]{@}{@},
  escapeinside={<@}{@>}
}

\lstdefinelanguage
   [x64]{Assembler}
   [x86masm]{Assembler}
   {morekeywords={CDQE,CQO,CMPSQ,CMPXCHG16B,JRCXZ,LODSQ,MOVSXD,
                  POPFQ,PUSHFQ,SCASQ,STOSQ,IRETQ,RDTSCP,SWAPGS,
                  rax,rdx,rcx,rbx,rsi,rdi,rsp,rbp,
                  r8,r8d,r8w,r8b,r9,r9d,r9w,r9b}}

\usepackage{listings}
\usepackage{fancyvrb}
\usepackage{color}
\definecolor{lightgray}{rgb}{.9,.9,.9}
\definecolor{darkgray}{rgb}{.4,.4,.4}
\definecolor{purple}{rgb}{0.65, 0.12, 0.82}
\definecolor{commentgreen}{RGB}{63,127,95}

\usepackage{xcolor}
\colorlet{myPurple}{blue!40!red}
\definecolor{myOrange}{RGB}{255,192,0}

\lstdefinelanguage{Solidity}{
  keywords={len,delete,int,void,payable, public, event, contract, typeof, new, true, false, catch, function, return, null, catch, switch, var, if, in, while, do, else, case, break,struct,const,socklen_t,sa_familty_t,char,sockaddr},
  keywordstyle=\color{violet}\bfseries,
  ndkeywords={class, export, boolean, throw, implements, import, this},
  ndkeywordstyle=\color{darkgray}\bfseries,
  identifierstyle=\color{black},
  sensitive=false,
  comment=[l]{//},
  escapeinside={(*@}{@*)},
  morecomment=[s]{/*}{*/},
  commentstyle=\color{commentgreen}\ttfamily,
  stringstyle=\color{red}\ttfamily,
  morestring=[b]',
  morestring=[b]"
}

\newcommand{\rnum}[1]{\uppercase\expandafter{\romannumeral #1\relax}}

\usepackage{xcolor}

\definecolor{pptbrown}{RGB}{132,60,12}
\definecolor{pptgreen}{RGB}{56,87,35}

\newcommand{\syz}{\textsc{Syzkaller}}
\newcommand{\ms}{\textsc{Moonshine}}
\newcommand{\tool}{\textsc{RLTrace}}

\begin{document}

\title{\tool: Synthesizing High-Quality System Call Traces for OS Fuzz Testing}

\author{Wei Chen\inst{1} \and
Huaijin Wang\inst{1} \and
Weixi Gu$^{*}$\inst{2}\and
Shuai Wang\thanks{Corresponding authors.}\inst{1}}
\authorrunning{Chen et al.}
\institute{Hong Kong University of Science and Technology, Hong Kong, China\\
\email{\{wchenbt, hwangdz, shuaiw\}@cse.ust.hk}\and
China Academy of Industrial Internet, Beijing, China\\
\email{guweixi@china-aii.com}}

\maketitle

\begin{abstract}
Securing operating system (OS) kernel is one central challenge in today's cyber
security landscape. The cutting-edge testing technique of OS kernel is software
fuzz testing. By mutating the program inputs with random variations for
iterations, fuzz testing aims to trigger program crashes and hangs caused by
potential bugs that can be abused by the inputs. To achieve high OS code
coverage, the de facto OS fuzzer typically composes \textit{system call traces}
as the input seed to mutate and to interact with OS kernels. Hence, quality and
diversity of the employed system call traces become the prominent factor to
decide the effectiveness of OS fuzzing. However, these system call traces to
date are generated with \textit{hand-coded rules}, or by analyzing
\textit{system call logs} of OS utility programs. Our observation shows that
such system call traces can only subsume common usage scenarios of OS system
calls, and likely omit hidden bugs.

In this research, we propose a deep reinforcement learning-based solution,
called \tool, to synthesize diverse and comprehensive system call traces as the
seed to fuzz OS kernels. During model training, the deep learning model
interacts with OS kernels and infers optimal system call traces w.r.t. our
learning goal --- maximizing kernel code coverage. Our evaluation shows that
\tool\ outperforms other seed generators by producing more comprehensive system
call traces, subsuming system call corner usage cases and subtle dependencies.
By feeding the de facto OS fuzzer, \syz, with system call traces synthesized by
\tool, we show that \syz\ can achieve higher code coverage for testing Linux
kernels. Furthermore, \tool\ found one vulnerability in the Linux kernel
(version 5.5-rc6), which is publicly unknown to the best of our knowledge by the
time of writing. We conclude the paper with discussions on the
limitations, tentative exploration of technical migration to other OS kernels,
and future directions of our work. We believe the proposed \tool\ can be a
promising solution to improve the reliability of OS fuzzing in various
scenarios, over different OS kernels, and for different reliability purposes.
\end{abstract}

\vspace{-10pt}
\section{Introduction}
\label{sec:introduction}
An operating system (OS) kernel usually contains millions lines of code, with
complex program structures, deep call hierarchies, and also stateful execution
models. Nowadays, OS-level vulnerabilities are gaining more and more attention,
not only because it is usually much more challenging to be detected, but also
because OS vulnerabilities, once being exploited, can lead to whole-system
security breaches with much more severe damages. To date, real-world
vulnerabilities has been constantly reported from OS kernels on various
computing platforms, including the multi-purpose computers (e.g., Windows and
Mac OS), mobile phones, and also embedded devices. Demonstrated by industrial
hackers, such vulnerabilities can often lead to severe threats to the financial
stability and public safety towards tremendous amounts of users in the real
world~\cite{linuxvul1,skybox2019vul}.

Software fuzz
testing performs vast mutation towards program inputs, exercises its underlying
functionalities and reveals vulnerabilities residing within the target software
that is difficult to find by traditional testing tools~\cite{10.5555/1404500}.
Despite its simplicity, fuzz testing outperforms many vulnerability detection
techniques due to its efficiency and robustness. So far, fuzz testing has helped
to detect tremendous amounts of defects from real-life applications, including
PDF readers, web browsers, and commonly-used mobile
apps~\cite{godefroid2008automated,godefroid2012sage,godefroid2008grammar}.

To fuzz an OS kernel, the primary strategy is to extensively mutate inputs of
the system-call interface, since the interface serves as the main points to
interact between the OS kernel and user-level
applications~\cite{syzkaller,trinity}. The state-of-the-art (SOTA) OS fuzzer takes OS
system call traces as the fuzzing seed, and extensively mutates values of system
call parameters to achieve high kernel code
coverage~\cite{pailoor2018moonshine,han2017imf}. This naturally solves the
problem to generate valid inputs for a system call, for instance, a legitimate
file descriptor for \texttt{write} can be created by first calling \texttt{open}
and returning a file descriptor. More importantly, OS kernels are
\textit{stateful} software, meaning that the coverage of invoking each system
call depends on the OS kernel state created by previously executed system calls.
Therefore, de facto OS fuzzers often take traces of system calls as the starting
point (i.e., fuzzing seeds) to bootstrap the campaign.

Existing research work mostly relies on ad-hoc approaches to generating valid
system call traces as OS fuzzer
seeds~\cite{syzkaller,trinity,pailoor2018moonshine,han2017imf}. For instance,
the de facto industry strength OS fuzzer, \syz, pre-defines thousands of
hand-coded rules to encode dependencies among different system calls (see
\S~\ref{subsec:os-kernel-fuzzing} on why ``dependencies'' are critical) and use
them to generate system call traces. A recent
work~\cite{pailoor2018moonshine} extracts system call traces from the system
call logs of OS utility programs. Despite the simplicity, our observation (see
\S~\ref{sec:motivation}) shows that system call traces generated from logs or
manually-written rules could only subsume some \textit{commonly-seen cases}.
Rarely-used system call traces may not be included, and even for the executed
system calls, many of the corner usage scenarios may not be covered as well.
Indeed, the performance of software fuzzing tools largely depends on the quality
and diversity of their input seeds~\cite{alexandre2014opt,wang2017skyfire}, and
as shown in our study, the quality of system call traces undoubtedly limits the
OS attack surface that can be tested, which further impedes OS fuzzers from
identifying real-world security flaws to a great extent.

In this research, we propose a unified and systematic solution, called \tool, to
synthesize high quality seeds to promote OS kernel fuzzing. \tool\ employs deep
reinforcement learning (RL), particularly Deep Q-Network (DQN), to synthesize
comprehensive sets of system call traces as the kernel fuzzing seeds.
\tool\ trains a DQN model to interact with the OS kernel and explore optimal
combinations of system calls. The kernel code coverage is used as the reward for
the learning process, and we feed our synthesized system call traces as the seed
of the SOTA OS fuzzer, \syz, to fuzz Linux kernels.
After training for 14.9 hours, \tool\ generates a set of 1,526 system call
traces to test the Linux kernel (ver. 5.5-rc6). We compare our synthesized
system call traces with the SOTA research in this field,
\ms~\cite{pailoor2018moonshine}, which leverages heavyweight static analysis
techniques to generate fuzzing seeds. Our evaluation shows promising findings:
42.0\% of traces synthesized by \tool\ overlap with outputs of \ms. Moreover,
manual study from the non-overlapped traces (58.0\%) shows that \tool\ can find
many corner cases where \ms\ is incapable of reasoning. Further inspection
reveals that the seed generated by \tool, without employing any pre-knowledge or
static analysis techniques, can extensively capture subtle dependencies among
system calls, outperforming seeds generated by \ms\ (for 300.0\%) and the
default seeds used by \syz\ (for 20.1\%). By fuzzing with seeds produced by
\tool for 24 hours, we successfully found one 0-day kernel vulnerability. 
Moreover, we illustrate the high generalizability of \tool\ by discussing the
migration to other (embedded) OS kernels. We show that \tool\ can be easily
migrated to fuzz other OS kernels, and we expect that the synthesized seeds can
achieve comparable performance as the seeds generated by \ms. {This
indicates the high potential of \tool\ in promoting OS kernel fuzzing in various
scenarios, over different OS kernels, and for different reliability purposes. We
leave it as our future work to extend \tool\ and demonstrate its high
generalizability on other OS kernels. We also discuss the limitations and future
directions of our work to paint a complete picture and the high potential of
\tool.}
In sum, we make the following contributions:

\begin{itemize}
  \item We introduce a \textit{new} focus to use a generative learning model to
    promote OS kernel fuzzing by synthesizing quality seeds --- system call
    traces. Our technique is unified and systematic, without relying on any
    manual-written rules or heavy-weight static analysis techniques.
  \item We build a practical tool named \tool\ in coordinating with the de facto
    industry OS fuzzer, \syz. Our throughout design and implementation enables
    the comprehensive testing of production Linux kernels.
  \item Our evaluation shows that \tool\ can outperform the SOTA
    fuzzing seed generators by generating more comprehensive system call traces,
    achieving higher code coverage, and unveiling more vulnerabilities.
  \item We present a case study to demonstrate the high generalizability
    of \tool\ by discussing the migration to other OS kernels. We show that
    \tool\ can be easily migrated to fuzz other OS kernels, and we expect that
    by extending \tool, we shall be able to achieve comparable performance and
    constantly uncover security defects of various properties on other OS
    kernels or platforms. We accordingly discuss the limitations and future
    directions of our work.
\end{itemize}

\begin{figure}
  \vspace{-20pt}
  \subfigure[Overview of OS fuzzers.]{
    \includegraphics[width=0.45\linewidth]{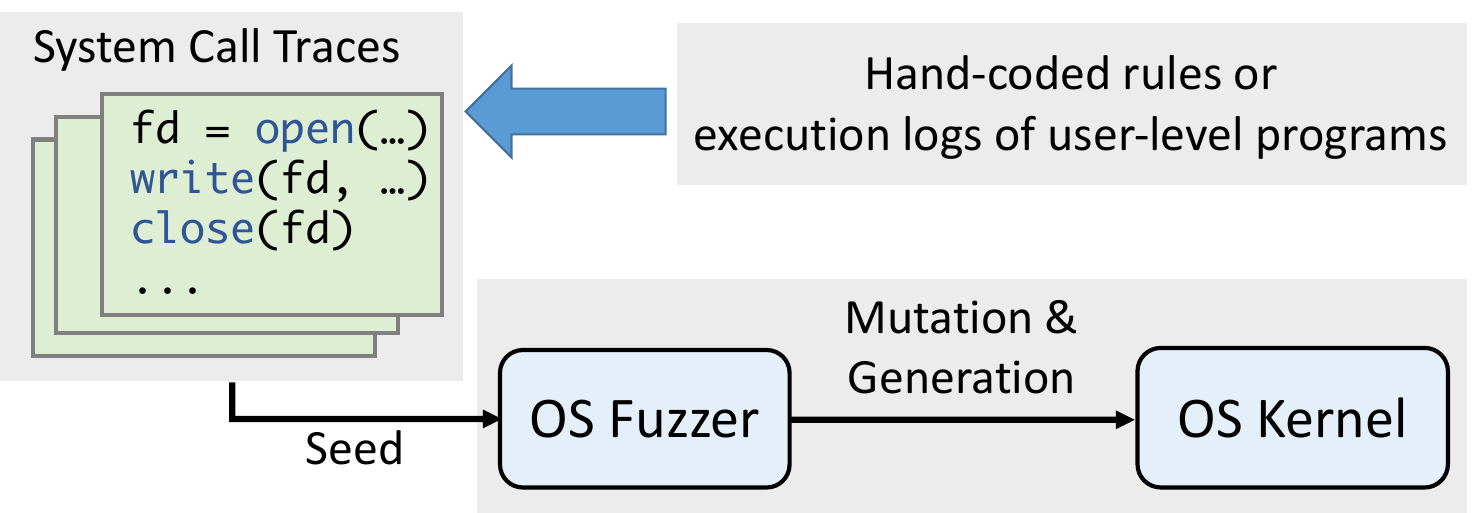}
    \label{fig:os-fuzzing}
  }
  \hfill
  \subfigure[Overview of \tool with learning.]{
    \includegraphics[width=0.45\linewidth]{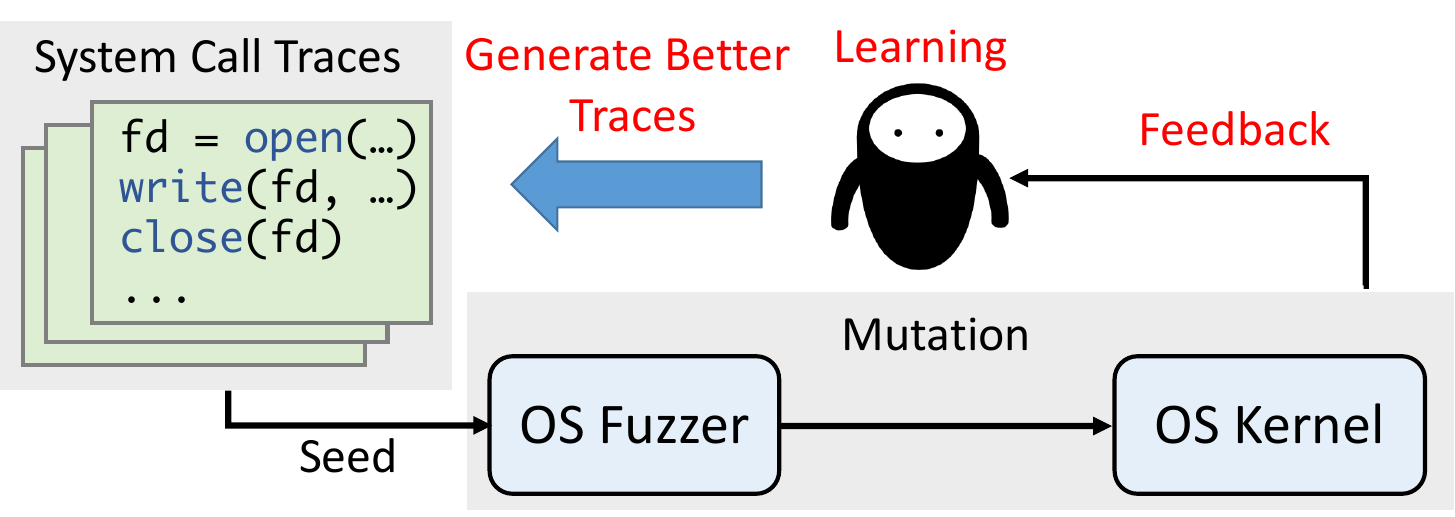}
    \label{fig:os-learning}
  }
  \vspace{-10pt}
  \caption{Overviews of OS fuzzers and \tool.}
  \vspace{-20pt}
\end{figure}

\section{Background}
\label{sec:background}
\subsection{Testing OS Kernels}
\label{subsec:os-kernel-fuzzing}
To secure OS kernels, the de facto technique is OS kernel \textit{fuzz
  testing}~\cite{pailoor2018moonshine,han2017imf}. The SOTA OS
fuzzers, \syz~\cite{syzkaller} and Trinity~\cite{trinity}, take a set of
\textit{system call traces} (each set is called a ``corpus'') as their seed
inputs for fuzzing. \F~\ref{fig:os-fuzzing} presents an overview of OS fuzzing
workflow. By feeding a corpus of system call traces into the OS fuzzer, the OS
fuzzer will vastly perturb the corpus (including fuzzing parameter values and
shuffling system calls on the trace) to interact with the OS kernel. Advanced OS
fuzzer like \syz\ can also generate new traces during the fuzzing campaign (see
\S~\ref{subsec:syz-instrument} on how \syz\ mutates and generates new traces).
Taking system call traces as the fuzzing inputs is intuitive. The execution of a
system call depends on the validity of its input parameters (e.g., a legitimate
file descriptor). In addition, internal kernel state created or changed by
previous system calls can also influence the execution of succeeding system
calls. Invoking a single system call without setting up the proper ``context''
would merely explore all the functional components of this system call.

\begin{figure*}[!htbp]
  \vspace{-20pt}
  \centering
  \includegraphics[width=1.00\linewidth]{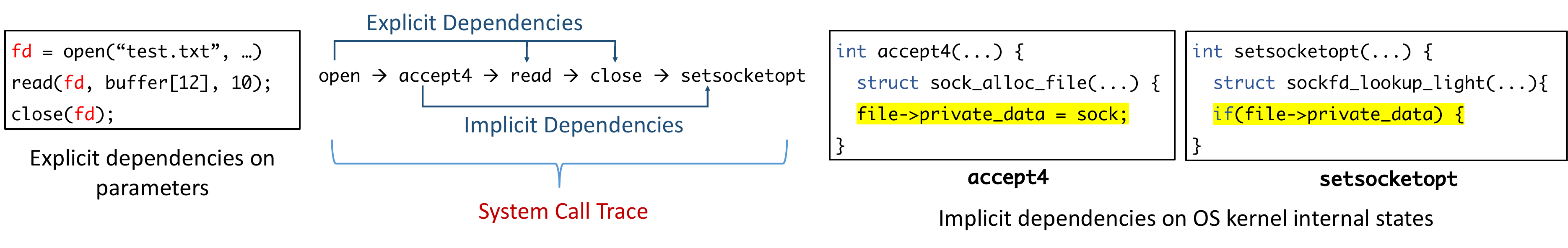}
  \vspace{-20pt}
  \caption{A sample system call trace used by the OS fuzzer, \syz. To achieve
    high coverage, both explicit and implicit dependencies need to be satisfied,
    which is quite challenging for existing rule-based or program analysis-based
    seed generators~\cite{syzkaller,pailoor2018moonshine}.}
  \label{fig:trace}
  \vspace{-10pt}
\end{figure*}

In fact, a common assumption shared by existing research and industry OS
fuzzers~\cite{pailoor2018moonshine,han2017imf,syzkaller,trinity} is that the
quality of fuzzing seeds heavily depends on the number of explicit and implicit
dependencies satisfied on each system call trace. \F~\ref{fig:trace}
presents a sample trace of five Linux system calls used as a high-quality seed.
System call \texttt{read} and \texttt{close} \textit{explicitly depend} on
system call \texttt{open}, since the output of \texttt{open} creates a valid
file descriptor which can be used as the input parameter of \texttt{read} and
\texttt{close}. As a result, simultaneously fuzzing \texttt{open},
\texttt{write}, and \texttt{close} can effectively cover functional components
of all these system calls~\cite{pailoor2018moonshine}, while merely fuzzing
\texttt{read} or \texttt{close} along may be trapped in the exception handling
routines for most of the time (since the inputs are usually invalid). More
importantly, a system call may \textit{implicitly depend} on another system
call, if the execution of one system call affects the execution of the other
system call via shared kernel data structures (i.e., OS \textit{internal
  states}), for example the execution of \texttt{accept4} and
\texttt{setsocketopt} both affect and depend on the socket's private data, and
by mutating the parameters of \texttt{accept4}, execution of
\texttt{setsocketopt} will cover different paths in the kernel (see the third
diagram in \F~\ref{fig:trace}), although their parameters do not depend on each
other explicitly.

While explicit dependencies can be summarized by analyzing the system call
documents, obtaining implicit dependencies among system calls, however, is very
difficult. Typical (closed-source) OS kernels are highly complex and conducting
precise program analysis to pinpoint such dependencies are very challenging, if
at all possible. Indeed, as shown in \F~\ref{fig:os-fuzzing}, existing OS
fuzzers derive system call traces with manually-written rules, or by analyzing
system call logs of OS utility programs to infer
dependencies. However, our investigation (\S~\ref{sec:motivation}) shows
that these ad-hoc approaches have limited comprehension and presumably miss
hidden defects in the kernel. In contrast, our research takes a learning
approach to synthesize diverse system call traces from
scratch. Our evaluation shows \tool\ successfully finds four times more
implicitly dependent system calls, without any manual efforts (see
evaluations reported in \T~\ref{tab:dependency}).

\begin{table}
\centering
\caption{Kernel data access dependencies of a Linux system call
  \texttt{pwritev}. We report that the state-of-the-art OS fuzzing seed
  generator (\ms~\cite{pailoor2018moonshine}) only covers
  \textcolor{red}{\texttt{open}}.}
  \vspace{-10pt}
\resizebox{1.0\textwidth}{!}{
		\begin{tabular}{|c|c|}
			\hline
			Total Number & System Call Names \\
			\hline
 			\multirow{3}{*}{27} &  \texttt{openat}; \texttt{mq\_open}; \texttt{epoll\_ctl}; \texttt{shmdt}; \texttt{epoll\_create1}; \texttt{mmap\_pgoff}; \texttt{fadvise64\_64}; \texttt{swapoff}; \texttt{acct} \\
 &  \texttt{shmctl}; \texttt{msync};  \texttt{flock};  \texttt{\textcolor{red}{open}};  \texttt{uselib};  \texttt{accept4};  \texttt{dup};  \texttt{setns};  \texttt{socketpair};  \texttt{remap\_file\_pages};  \texttt{dup3} \\
  &   \texttt{shmat}; \texttt{socket}; \texttt{open\_by\_handle\_at}; \texttt{memfd\_create}; \texttt{pipe2}; \texttt{eventfd2}; \texttt{perf\_event\_open} \\
 			\hline
		\end{tabular}
}
\label{tab:data}
\vspace{-10pt}
\end{table}

\subsection{Deep Reinforcement Learning (DRL)}
\label{subsec:dqn}
\tool\ is built on top of a deep reinforcement learning (DRL) model to
synthesize quality system call traces. RL is a framework that trains an agent's
behavior by interacting with the surrounding environment. During the learning
process, the agent observes the environment and performs actions accordingly.
For each step of interaction, the agent earns some rewards from the environment,
and usually the system goes through a state transition as well. During the
overall time of learning, the agent gradually learns to maximize its cumulative
reward (i.e., a long-term objective).

We formulate a typical RL process related to the presentation given
in~\cite{szepesvari2010algorithms}, where the action of the agent can be viewed
as a stochastic process. In particular, the Markov decision procedure
is a triple $\mathcal{M} = (\mathcal{X}, \mathcal{A}, \mathcal{P})$, where
$\mathcal{X}$ denotes the set of states in the environment, $\mathcal{A}$ is the
set of actions an agent can take, and $\mathcal{P}$ represents the
\textit{transition probability kernel}, which assigns a probabilistic value
denoted as $\mathcal{P}(\cdot|x,a)$ for each state-action pair (x, a) $\in
\mathcal{X} \times \mathcal{A}$.
For each reward $U \in \mathbb{R}$,
$\mathcal{P}(U|x,a)$ gives probability such that performing action $a$ at state
$x$ engenders the system to transition from $x$ into $y \in \mathcal{X}$ and
returns reward value $U \in \mathbb{R}$.
During the stochastic process $(x_{t+1},r_{t+1}) \sim
\mathcal{P}(\cdot|x_{t},a_{t})$, the goal of the agent is to choose a sequence of
actions to maximizes the expected cumulative rewards
$\mathcal{R} = \sum_{t=0}^{\infty}\gamma^{t}R_{t+1}$,
where $\gamma \in (0, 1)$ is discount factor.

The deep Q-network (DQN) technique is a specific approach for training
a model to select optimal sequences of actions.
It has been broadly used in solving real-world
challenges and achieved prominent success, including playing strategy board
games~\cite{silver2016mastering} and video games~\cite{mnih2013playing}. In this
research, we show that DQN can be leveraged to synthesize quality OS fuzzing
seeds and outperform existing rule-based or log-based seed generators.

\vspace{-5pt}
\section{Limitation of De Facto OS Fuzzers}
\label{sec:motivation}
The de facto industry strength OS fuzzer, \syz,
implements thousands of \textit{manually-written} rules to summarize potential
dependencies among system calls and generate system call traces (i.e., default
seeds shipped with \syz). Nevertheless, it has been pointed
out~\cite{pailoor2018moonshine} (and also consistently reported in our
evaluation; see \S~\ref{sec:evaluation}) that such rule-based approach cannot
achieve highly effective fuzz testing. The reason is that many of
its generated traces are \textit{lengthy and repetitive}. From a holistic
view, while having a large number of system calls on each trace
intuitively improve the ``diversity'' and coverage, an unforeseen drawback is
that within a given amount of time, fuzzer would perform less throughout
exploration for each individual system call (too many system calls on a trace),
thus scaling down the coverage.

In contrast, the SOTA research, \ms~\cite{pailoor2018moonshine},
generates system call traces by analyzing system call logs of OS utility
programs. By further performing OS kernel dependency analysis, this work detects
dependencies across different system calls to ``distills'' system call logs;
system call traces will primarily retain system calls that depend on each other.
It is reported that \ms\ can largely promote the fuzzing efficiency compared to
the vanilla seeds of \syz~\cite{pailoor2018moonshine}. Intuitively, by
\textit{decreasing} the number of system calls present on a trace and only
focusing on system calls dependent on each other, the fuzzer can allocate more
time to mutate inputs of each system call, and likely increase the
code coverage. Nevertheless, our preliminary study shows that system call traces
simplified from program execution logs become \textit{less comprehensive} and
insufficient to cover the diverse set of Linux kernel system calls.

We collecte system call traces covered by \ms\ and compared them with the whole
set of Linux system calls. Linux kernel (version 5.5-rc6) has 407 system calls,
and the OS fuzzing tool, \syz, supports 331 system calls. We
report that out of these 331 system calls, \ms\ can only cover 180 system
calls (53.9\%; see \T~\ref{tab:system-call}) since certain system calls are never
used by selected OS utility programs or are trimmed off after its
dependency analysis. Moreover, \ms\ can hardly consider all usage scenarios of
a system call and rarely-used system calls may not be included since
it depends on selected programs, which undoubtedly limits the 
OS attack surface that can be tested.

\vspace{-5pt}
\section{Design}
\label{sec:design}
\F~\ref{fig:os-learning} depicts the overview of the proposed technique. Instead
of pulling out system call traces from execution logs or some manually-written
rules, we synthesize system call traces from scratch with the guidance of
learning-based methods and with the learning goal of achieving high code
coverage. The synthesized traces would form a diverse and comprehensive basis to
explore OS kernels, by smoothly taking rarely-used system calls and different
system call execution contexts into account, as long as they notably increase
code coverage.

Inspired by recent advances in RL-based program analysis and
testing~\cite{si2018nips,godefroid2017learn,gupta2019deep}, where an agent is
trained to learn good policies by trial-and-error, we leverage RL models to
synthesize system call traces and solve this demanding problem. \tool\ is
constructed as a DQN with two fully connected layer with non-linear activation
function \texttt{relu}. Each hidden layer contains 512 hidden units. We encode
system call traces as a practical learning representation
(\S~\ref{subsec:state}), and our agent is trained to continuously perturb system
calls on a trace (\S~\ref{subsec:action-space}). The code coverage will be used
as the learning reward to train the agent (\S~\ref{subsec:search-reward}). For
each episode, the model training forms an iterative process until our cumulative
reward becomes higher than a predefined threshold $T_{1}$, or becomes lower than
another threshold $T_{2}$. Parameters $T_{1}$ and $T_{2}$ can be configured by
users. We harvest optimal traces and pack them into a seed file (i.e., named
``corpus'') for use.

\noindent \textbf{Application Scope.}~\tool\ is evaluated on widely-used Linux
OS kernels. Although the source code is available, we treat the OS kernel as a
``black-box'' (no need for source code). Hence, the proposed techniques can be
used during in-house development where source code is available, and also
smoothly employed to test \textit{closed-source} OS kernels (e.g, Windows or Mac
OS). In contrast, one SOTA seed generator, \ms, performs heavy-weight static
dependency analysis on the source code of the Linux kernel. One may
question if \tool, to some extent, is only applicable to mainstream OS kernerls
(e.g., Linux) which are fully and clearly documented. We however anticipate that
\tool\ can be seamlessly integrated to test commercial, embedded OS kernels even
if the APIs documents are not fully disclosed (e.g., due to commercial
confidentiality or IP protection). We believe there is no major technical
challenges with the enhancement of modern learning techniques like transfer
learning or active learning. See our discussions on extension and future
directions in \S~\ref{sec:discussion}.

\subsection{State}
\label{subsec:state}

State is a basic element in formulating a RL learning procedure.
We define a state is a trace of OS system calls $(f_1, f_2, \ldots, f_L)$ where
$L$ is the length of the trace (see \S~\ref{sec:tool}).
We encode the system call with one-hot embedding. Then, a system call trace
is treated as a set of system calls, which is encoded with the sum of all
system call embeddings.
The reason to not adopt sequential embedding methods (e.g., LSTM~\cite{gers1999learning})
is that \syz\ indeed \textit{shuffles}
each system call trace during fuzzing. In other words, it is sufficient to only
use the multi-hot embedding (no need to preserve the ``order'' with sequential
embedding methods).

Typical RL process takes a considerable number of episodes to train an agent:
each episode denotes a sequence of state changes starting from the initial state
and ending at a terminal state (see \S~\ref{subsec:search-reward} for terminal
state definition). In this research, we randomly generate the initial state as
the starting point of each episode. Also, note that the encoding is only fed to
the agent for learning within each episode. We translate each numeric encoding
in a state back to its corresponding OS system call before fuzzing the OS
kernel.

Deciding an optimal and adaptive length of each trace is difficult. As discussed in
\S~\ref{sec:motivation}, deciding optimal length forms a dilemma: by decreasing
the number of system calls present on the trace, the fuzzer allocates more time
to mutate inputs of each system call. However, succinct trace may only subsume
limited system calls and prone to missing a large portion of OS interfaces,
exposing negative effects on code coverage. Similarly, lengthy traces (e.g., the
default seed of \syz) possess more system calls, but can allocate less time to
mutate each individual call. Given the general challenge to infer an optimal
length, we resort to launch empirical studies and find out that length $L=5$
usually leads to favorable coverage (see
\S~\ref{subsec:length}).

\subsection{Action}
\label{subsec:action-space}
We now define all the actions that an agent can perform toward the state during
the learning. Given a state $(f_{1}, f_{2}, \ldots, f_{L})$ which consists of
$L$ system calls, we mimic how a human agent could take actions and perturb the
trace. Overall, the agent first navigates within the trace, flags certain
$f_{i}$ where $i \in [1, L]$, and then updates the state by changing $f_i$ to
some other system call $f'_i$. While a ``random'' navigation could provide
maximal flexibility by selecting one arbitrary element $f_{i}$ within the state
to perturb, the search space is indeed quite large: $L\times N$ where $N$ is the
total number of system calls that can be tested (331 for Linux kernel 5.5-rc6).

To practically reduce the search space, our agent starts from the first element
$f_{1}$ to mutate, and each learning step only moves one element forward. When
it reaches the end, it will re-start from the first element again. This way, our
agent picks only one system call $f_{i}$ each learning step and replaces it with
a predicted $f'_{i}$. Note that our agent can also retain the current state, as
long as $f_{i}$ equals to $f'_{i}$. The search space of our agent is reduced
into $N$. Evaluation shows that the model training is efficient in practice.
Although there are still considerable states to explore (since $N$ is still
large), deep Q-networks have been shown to handle large state spaces
efficiently, as we will show in \S~\ref{sec:evaluation}.

\subsection{Reward}
\label{subsec:search-reward}
The reward function is the key to RL frameworks. The central assumption behind
the SOTA coverage-based fuzzer is that coverage increase indicates a higher
possibility of detecting new vulnerabilities. Therefore, to construct the search
witness, we take the coverage increase into account. The OS fuzzer employed in
this research provides basic block coverage for each individual system call on
the trace for use. Let $c_{1} c_{2} \dots c_{i} \dots c_{L}$ be the code
coverage of individual system calls on the trace $s = (f_{1} f_{2} \dots f_{i}
\dots f_{L})$. Suppose by replacing $f_{i}$ with $f'_{i}$, the produced new
system call trace is $s' = (f_{1} f_{2} \dots f'_{i} \dots f_{L})$. Then, the
reward function w.r.t. this action $f_{i} \rightarrow f_{i'}$ is formulated as,
$ R = \frac{\sum_{i=1}^{L}\log \frac{c'_{i}}{c_{i}}}{L}$, where $c'_{i}$ are the
code coverage of individual system calls on our new trace. The learning reward
is a positive number, in case a higher code coverage is achieved by the new
trace. Nevertheless, we penalize coverage decrease by computing and assigning a
negative reward to the agent.

Overall, the coverage feedback depends on the entire system call trace
synthesized so far rather than on the very last system call being picked. In
other words, our formulation indeed takes long-term rewards into consideration,
which progressively infers the optimal system call traces.

\noindent \textbf{Hyperparameters.}~For long-term reward harvesting, we use a
discount rate of 0.9. Our learning rate is 0.01. We follow the common practice
to adopt a $\epsilon$-greedy policy with $\epsilon$ decayed from 0.95 to 0.
$\epsilon$ will be fixed at 0 thereafter. The agent will select the predicted
optimal action with the probability of $1-\epsilon$, and explores random actions
with the probability of $\epsilon$. Hence, the training starts by focusing on
random explorations and gradually converge to optimal decisions. Overall, while
we follow common and standard practice to decide model hyperparameters and
settings, evaluation results already report promising findings.

\noindent \textbf{Terminal State.}~In the context of RL, ``episode'' defines a
sequence of state transitions which ends with terminal state. A RL model
training usually takes hundreds of episodes until saturation.
We define the terminal state such that the cumulative reward is greater than a
threshold $T_{1}$ (10.0 in our current implementation). The current episode will
also be terminated, if the cumulative reward is lower than another threshold
$T_{2}$ (-5.0 in our current implementation), indicating that there should be
few chances we can find an optimal trace during this episode. Hence, we
terminate the current episode. At the end of an episode, we archive the
synthesized optimal trace for reuse. Archived traces will be packed into a
corpus and fed to the OS fuzzer as its seed.

\section{Implementation}
\label{sec:tool}
The RL learning framework is built on top of Tensorflow (ver. 1.14.0), written
in Python (about 500 lines of code). We also instrumented the OS fuzzer,
\syz\ (see \S~\ref{subsec:syz-instrument}). This patch is written in Go, with
approximate 300 lines of code.

\subsection{\syz\ Instrumentation}
\label{subsec:syz-instrument}
The de facto OS fuzzer used in this research, \syz, is widely used in testing
real-world OS kernels and has been constantly finding (security-critical) bugs.
Given a seed generated by packing system call traces, \syz\ performs four
mutation strategies including 1) mutating system call parameters, 2) shuffling
system calls on a trace, 3) removing system calls on a trace or adding extra
calls, and 4) generating new system call traces from hand-written rules and
templates. In general, the first three strategies primarily depend on the
quality of seeds, while the last one implements a carefully crafted
``generation'' strategy to synthesize new inputs during the long run. Although
the standard \syz\ mingles all four mutation strategies together, during model
training, we instrument \syz\ and only enable the first three strategies to
better reflect quality of synthesized seeds. Similarly for the fuzzing
evaluation, we measure the seed quality by only using the first three strategies
(\S~\ref{subsec:coverage-comparison}). We also resort to the default setting of
\syz\ with all strategies enabled to mimic the ``real-world'' usage in the
evaluation (\S~\ref{subsec:fuzzing-generation}).

The standard workflow of \syz\ requires to re-boot the tested OS kernel (hosted
in a virtual machine instance) every time before fuzzing. Hence for every
learning step, we need to terminate and reboot the VM instance, exposing high
cost to model training. To optimize the procedure, we instrument \syz\ by adding
an agent module. After booting the VM instance for the first time, the agent
listens for requests from \tool\ and forwards synthesized traces to the fuzzer
module of \syz. In this way, the VM instance will be booted for only once during
the entire training. This instrumentation reduces the training time from 67.4
CPU hours to 14.9 CPU hours (see \S~\ref{subsec:learning-setup} for model
training).

\begin{figure}
  \vspace{-10pt}
  \subfigure[Kernel code coverage w.r.t. different trace lengths.]{
    \includegraphics[width=0.45\linewidth]{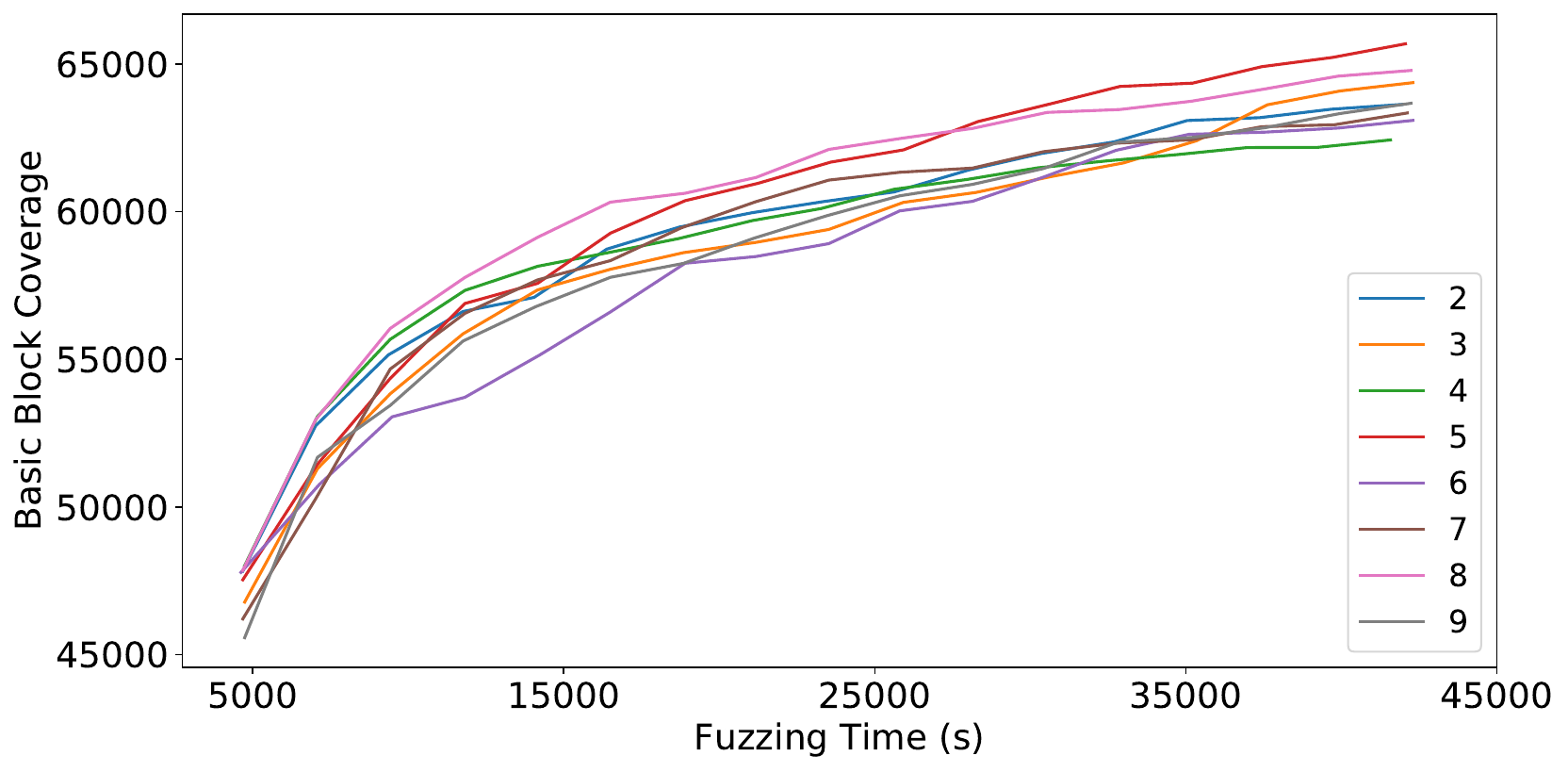}
    \label{fig:length}
  }
  \hfill
  \subfigure[Loss function decrease over episodes. We present overall 80k steps
  corresponding to 480 episodes trained in this evaluation.]{
    \includegraphics[width=0.45\linewidth]{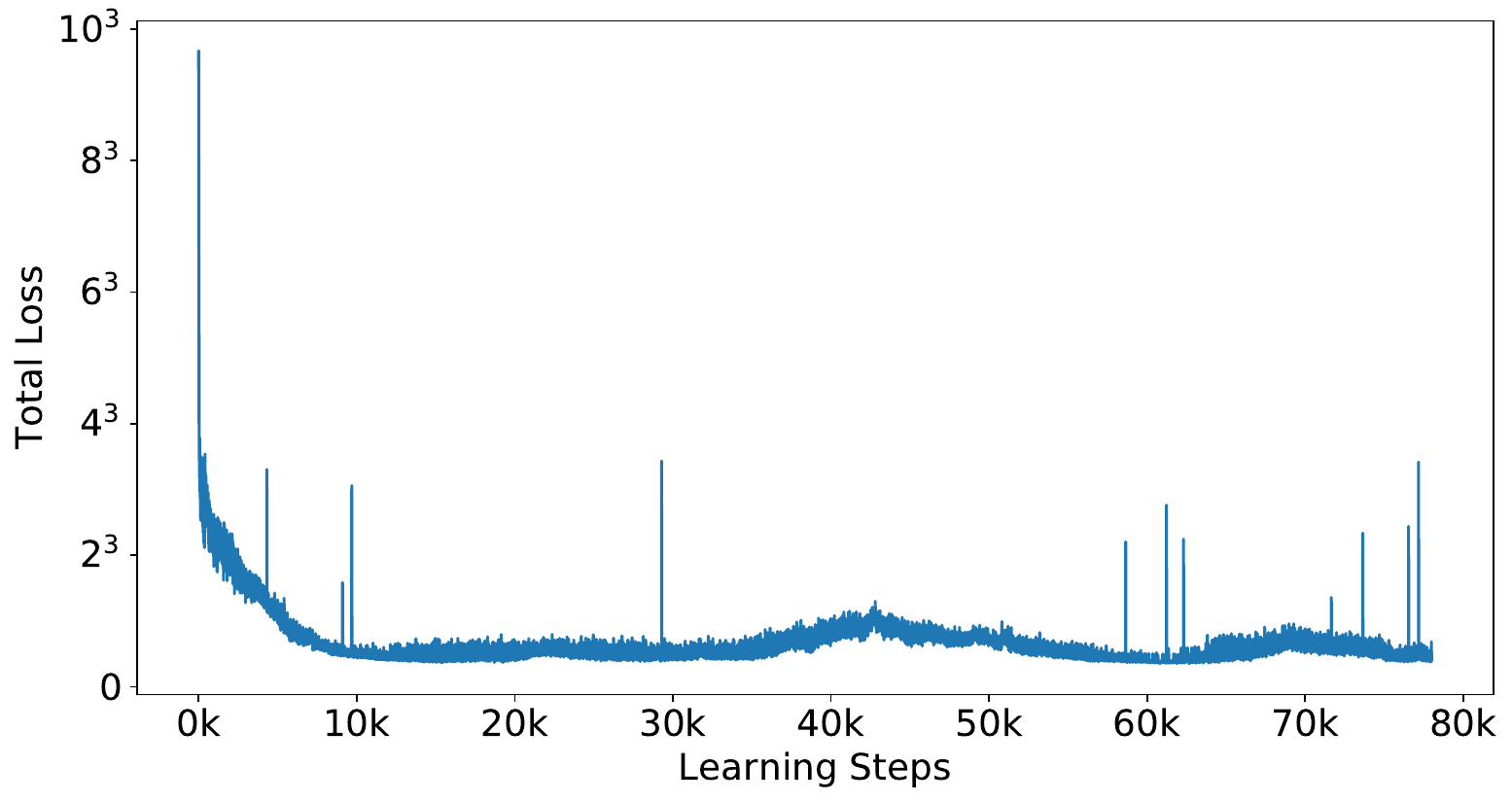}
    \label{fig:reward}
  }
  \vspace{-10pt}
  \caption{Kernal code coverage and loss function curves.}
\end{figure}

\subsection{Decide the Length of System Call Trace}
\label{subsec:length}
To decide the length of an optimal system call trace, we launch empirical
studies to explore the change of length with respect to their corresponding
coverage data. In general, our observation shows that too lengthy traces can
notably slow down the fuzzer, and therefore are not desired. Hence, we
empirically decide to vary the trace length from two to nine and record the
corresponding basic block coverage.

Let $N$ be the number of total system calls we decide to use in this study, we
start by randomly constructing $S_{2}$ system call traces with a fixed length as
two, such that $S_{2}\times 2 \approx N$. We then feed \syz\ with this set of
traces, fuzz the Linux kernel, and record the coverage increase. We then
randomly construct $S_{3}$ traces of three system calls (to present a fair
comparison, here $S_{3}\times 3 \approx N$) and re-launch the whole experiments
until we have tested $S_{9}$ traces of nine system calls (again, $S_{9}\times 9
\approx N$). For this study, $N$ is 7,679 which equals to the total number of
system calls used in the seed generated by \ms\ (see relevant information in
\S~\ref{sec:evaluation}). \F~\ref{fig:length} reports the evaluation results in
terms of basic block coverage increase. We report that traces with five system
calls can outperform other settings with sufficient time of fuzzing (after about
27,000 seconds).
Given this empirical observation, the implementation adopts five as the length
of each synthesized system call traces. Indeed, we report that empirical results
revealed at this step is essentially consistent with \ms: the average length of
traces generated by \ms\ is 5.2.

\vspace{-5pt}
\section{Evaluation}
\label{sec:evaluation}
We evaluate the proposed technique in terms of its ability to promote real-world
OS fuzz testing. As aforementioned, we feed the synthesized system call traces
into an industrial-strength OS fuzzer, \syz~\cite{syzkaller}, to fuzz the Linux
kernel. We use Linux kernel version 5.5-rc6 (released 12 January 2020)
for the evaluation unless stated otherwise.
To evaluate the effectiveness of \tool, we compare the outputs of \tool\ with
the SOTA OS fuzzing input generator,
\ms~\cite{pailoor2018moonshine}, and also the default seeds of \syz\ generated
by hand-written rules. For the ease of presentation, we use $S_{rl}$ and
$S_{moon}$ to represent seeds generated by \tool\ and \ms, and $S_{def}$ to
represent the default seeds of \syz.

The empirical study in \S~\ref{subsec:length} decides the length of each system
call trace as five. To provide a fair comparison, we first compute the total
number of system calls from $S_{moon}$ ($S_{moon}$ is shared by the paper
author): we report that from 525 traces in $S_{moon}$, 7,679 system calls are
included. Hence, we decide to configure \tool\ and generate 1,526 traces. Recall
as introduced in \S~\ref{subsec:syz-instrument}, \syz\ implements mutation
strategies to extend certain traces with extra system calls. We observe that
that when feeding these 1,526 traces into \syz, \syz\ indeed extends certain
traces with in total 47 extra system calls (e.g., when detecting
\texttt{timespec}, \syz\ will add \texttt{clock\_gettime} ahead of
\texttt{timespec}). In short, the total system call numbers in $S_{rl}$ is 7,677
($1526\times5 + 47$). Also, we confirm that \textit{no} extra system calls need
to be inserted into $S_{moon}$ when fed to \syz; this is reasonable since
$S_{moon}$ is derived from system call logs, where real programs are generally
required to ``compensate'' extra system calls.

\syz\ leverages hand-written rules to generate a large amount of lengthy and
repetitive traces: we randomly select 1,027 traces from $S_{def}$ that also
include 7,679 system calls in total (no extra system calls are added as well).
Overall, while $S_{rl}$ and its two competitors have different number of traces,
the total number of system calls are (almost) identical, qualifying a fair
comparison.

\noindent \textbf{A Fair Comparison.}~It is easy to see that \tool\ can smoothly
synthesize more traces with little cost, by simply taking more episodes. In
contrast, \ms\ and \syz\ are bounded by the availability of quality OS test
suites or expert efforts. We consider this actually highlights the
conceptual-level \textit{difference} and \textit{advantage} of \tool.
Overall, we would like to emphasize that our evaluation in the rest of this
section presents a fair comparison which indeed \textbf{undermines} the full
potential of \tool.

\vspace{-5pt}
\subsection{Model Training}
\label{subsec:learning-setup}

We first report the model training results. The training was conducted on a
server machine with an Intel Xeon E5-2680 v4 CPU at 2.40GHz and 256GB of memory.
The machine runs Ubuntu 18.04. The training takes in total 14.9 CPU hours for
480 episodes. As reported in \F~\ref{fig:reward}, the loss function keeps
decreasing until reaching low total loss scores (after about 9.5K steps). We
interpret the results as promising; the trained model can be progressively
improved to a good extent and find optimal system call traces more rapidly along
the training. We then randomly select 1,526 optimal traces and pack them into a
seed (i.e., ``corpus''). Further studies on the seed quality
(\S~\ref{subsec:seed-comparison}) and code coverage (\S~\ref{subsec:fuzzing})
will be conducted on this seed.
\vspace{-5pt}
\subsection{Exploring System Call Trace Quality}
\label{subsec:seed-comparison}
\vspace{-5pt}
\subsubsection{Cross Comparison}
\label{subsubsec:cross-comparison}
We now take a close look at $S_{rl}$ and compare it with $S_{moon}$ and
$S_{def}$. To this end, we first count and compare the number of unique system
calls that are included in these three seeds.

\begin{table}
  \vspace{-10pt}
  \centering
  \scriptsize
  \caption{System call coverage and explicit/implicit dependencies comparison.}
  \vspace{-10pt}
  \label{tab:system-call}
  \label{tab:dependency}
  \resizebox{0.90\linewidth}{!}{
    \begin{tabular}{l|c|c|c}
      \hline
      \textbf{Seed}          & \textbf{\#Covered Unique System Calls} & \textbf{Explicit Dependency} & \textbf{Implicit Dependency}\\
      \hline
      $S_{rl}$                 & 291 & 423  & 376 \\
      $S_{moon}$               & 180 & 247  & 94  \\
      $S_{def}$                & 115 & 775  & 313 \\
      \hline
    \end{tabular}
  }
  \vspace{-10pt}
\end{table}

\T~\ref{tab:system-call} reports the comparison results which is encouraging.
Out of in total 331 Linux system calls, 1,526 traces synthesized by \tool\ cover
291 unique system calls. In contrast, $S_{moon}$ has a low coverage: from in
total 525 system call traces subsumed in $S_{moon}$, only 180 unique system
calls are covered. $S_{def}$ yields an even worse coverage: 115 unique system
calls are included.

\begin{figure}
  \vspace{-25pt}
  \subfigure[Comparison of system call usages on the trace.]{
    \includegraphics[width=0.45\linewidth]{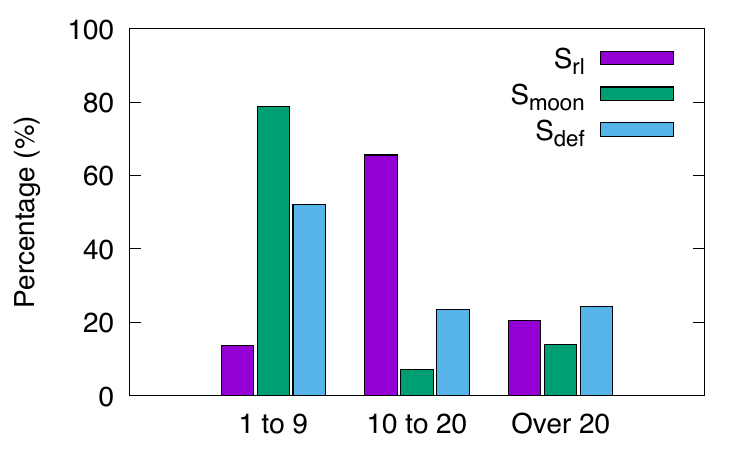}
    \label{fig:usage-bar}
  }
  \hfill
  \subfigure[Agreement among system call traces.]{
    \includegraphics[width=0.35\linewidth]{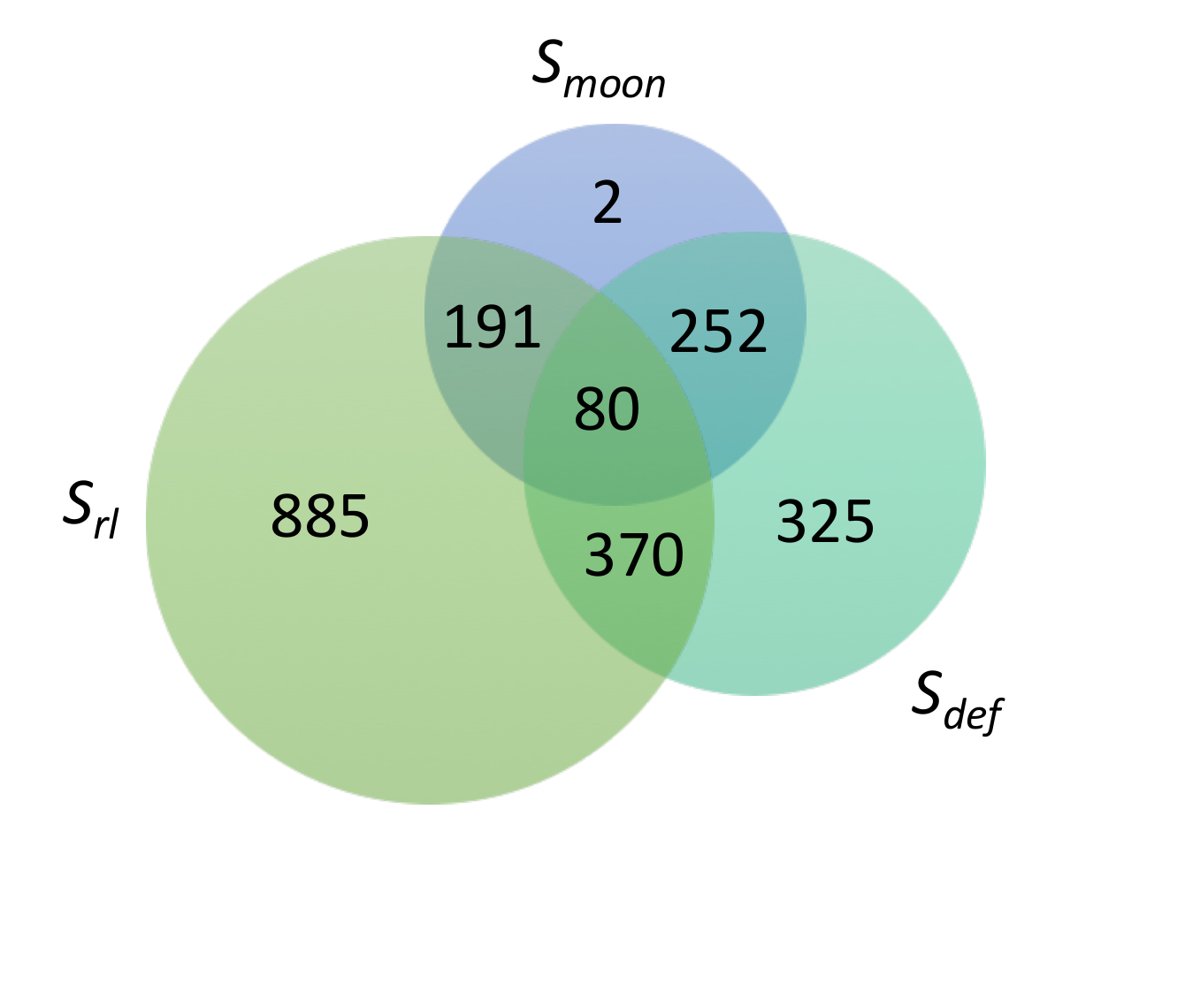}
    \label{fig:venn}
  }
  \vspace{-10pt}
  \caption{Analysis of call trace quality.}
  \vspace{-10pt}
\end{figure}

\F~\ref{fig:usage-bar} reports and compares the usage of each system call in
these three seeds. In particular, we count how many different traces a system
call can be found from. As aforementioned, different traces intuitively denote
various ``execution contexts.'' We are anticipating to systematically explore a
system call, if more contexts can be provided.
As reported in \F~\ref{fig:usage-bar}, 78.9\% of system calls are used for only
less than ten times by $S_{moon}$. Similar trending can also be observed from
$S_{def}$. In contrast, the output of \tool\ domains the second range: majority
system calls are used for over ten times. Overall, we interpret that
\T~\ref{tab:system-call} and \F~\ref{fig:usage-bar} demonstrate promising and
intuitive findings. Enabled by the systematic and in-depth exploration in \tool,
less commonly used system calls and diverse contexts can be taken into account
as long as they reasonably contribute to code coverage.

We also measured the agreement of three seeds by counting the number of
equivalent traces. We relax the notion of trace ``equivalence'' by entailing the
partial order of two traces (note that traces are shuffled within \syz\ and
therefore is reasonable to treat as ``sets'' without considering orders): $t
\doteq t' \leftrightarrow t \subseteq t' \lor t' \subseteq t$.

\F~\ref{fig:venn} reports the analysis results. $S_{rl}$ and $S_{moon}$ agree on
271 traces, while $S_{rl}$ and $S_{def}$ agree on 450. $S_{moon}$ and $S_{def}$
have 332 agreements (63.2\% of all traces in $S_{moon}$; highest in terms of
percentage). Overall, we interpret the results as promising: we show that
considerable amount of system call traces (641; 42.0\% of all traces in
$S_{rl}$) can be synthesized without employing heavy-weight program analysis or
hand-coded rules. Moreover, the 58.0\% disagreement, to some extent, shows that
\tool\ enables the construction of more diverse system call traces enable the
coverage of corner cases. Accordingly, we now present case studies on system
call traces generated by \tool.

\noindent \textbf{Case Study.}~Our study shows that \texttt{pwritev} implicitly
depends on 27 system calls of the Linux kernel. $S_{rl}$ consists of 10 traces
containing \texttt{pwritev}, and these 10 traces cover three unique system calls
that \texttt{pwritev} implicitly depends on. In contrast, as noted in
\S~\ref{sec:motivation}, \texttt{pwritev} and only one of its implicitly
dependent system call can be found from $S_{moon}$, and this particular case is
subsumed by $S_{rl}$ as well. Consider the system call trace below:
\vspace{-2pt}
\begin{lstlisting}
  lseek <@$\rightarrow$@> <@\textcolor{red}{openat}@> <@$\rightarrow$@> getxattr <@$\rightarrow$@> chmod <@$\rightarrow$@> pwritev
\end{lstlisting}
\vspace{-2pt}
\noindent where the system call \texttt{openat} and \texttt{pwritev} implicitly
depend on each other (i.e., they access and depend on the same kernel data
structure). We report that this system call trace can be found in $S_{rl}$, but
is not included in $S_{moon}$. In other words, kernel code coverage derived from
this implicit dependency will presumably not be revealed by \ms.

Similarly, we find another system call \texttt{fchown} implicitly depends on 97
system calls of the kernel. \tool\ generates 32 traces containing this system
calls, covering 15 unique system calls that \texttt{fchown} implicitly depends
on. For instance:
\vspace{-2pt}
\begin{lstlisting}
  <@\textcolor{red}{pipe2}@> <@$\rightarrow$@> getresuid <@$\rightarrow$@> fchown <@$\rightarrow$@> getresuid <@$\rightarrow$@> getpgrp
\end{lstlisting}
\vspace{-2pt}
\noindent where the system call \texttt{pipe2} and \texttt{fchown} implicitly
depend on each other. In contrast, we report that \ms\ only identifies
\texttt{one} implicitly dependent system call for \texttt{fchown}. Further
quantitative data regarding explicit/implicit dependencies is given in
\S~\ref{subsubsec:dependency}.

\vspace{-5pt}
\subsubsection{Dependency Analysis}
\label{subsubsec:dependency}
As discussed in \S~\ref{subsec:os-kernel-fuzzing}, OS fuzz testing leverages
system call traces of good quality, and therefore, aims at satisfying both
explicit and implicit dependencies and achieving high code coverage.
Nevertheless, directly analyzing dependencies could be challenging. We now
measure the quality of system traces, in terms of how they subsume explicit and
implicit dependencies.

\noindent \textbf{Explicit Dependency.}~Explicit dependencies denote system call
parameter and return value-level dependencies. To measure the performance, we
collect all the explicit dependencies of each system call. \syz\ provides a data
structure named \texttt{target} which can be parsed to acquire such information.
The summarized explicit dependencies (i.e., pairs of system calls; in total
4,429) deem a ``ground truth'' dataset, and we measure three seeds w.r.t. this
ground truth. The evaluation results are reported in the second column of
\T~\ref{tab:dependency}.

Overall, enabled by thousands of manually-written rules which extensively encode
system call dependencies among parameters and return values, $S_{def}$ largely
outperforms the other two seeds by recovering more explicitly dependent system
calls. \ms\ analyzes execution logs of OS utility programs to gather system call
traces. Real-world programs must satisfy these explicit dependencies to function
properly. Nevertheless, \tool\ still demonstrates encouraging results, by
inferring considerable explicit dependencies \textit{from scratch} and
outperform $S_{moon}$ (finding 176 more explicit dependencies). Envisioning the
necessity and opportunity of improving \tool\ at this step, we present
discussions in \S~\ref{sec:discussion}.

\noindent \textbf{Implicit Dependency.}~\ms\ releases a dataset to summarize
implicit dependencies among system calls, which is gathered by performing static
dependency analysis toward Linux kernel. Similarly, we reuse this dataset (in
total 9,891 pairs of system calls) to measure three seeds. Performing static
dependency analysis toward complex system software like Linux kernel is unlikely
to yield accurate results. Nevertheless, by measuring dependency recovery
regarding the same baseline, this is still an ``apple-to-apple'' comparison.

The third column of \T~\ref{tab:dependency} reports \textit{highly promising}
results. $S_{rl}$ notably outperforms its competitors, by finding more
implicitly-dependent system calls without adopting any hand-coded rules or
static analysis. $S_{def}$ primarily encodes explicit dependencies: implicit
dependencies are hard to be identified with only manual efforts.
Careful readers may wonder about the performance of \ms\ (since the ``ground
truth'' dataset is even provided by \ms). To clarify potential confusions:
\ms\ performs whole kernel static analysis to collect implicit dependencies. It
also performs execution trace-based analysis to collect system call traces. In
short, a system call trace will be kept if it matches certain implicit
dependencies collected by the static analysis. Since not all ``implicitly
dependent'' system calls will appear on execution traces, $S_{moon}$ does not
indeed perform well.

\vspace{-5pt}
\subsection{Fuzzing Evaluation}
\label{subsec:fuzzing}
\vspace{-5pt}
\subsubsection{Fuzzing without Runtime Trace Generation Enabled}
\label{subsec:coverage-comparison}
We feed \syz\ with three seeds to fuzz the Linux kernel 5.5-rc6. To faithfully
explore the quality of the generated seeds, we disable the ``generation''
strategy of \syz. Recall \S~\ref{subsec:syz-instrument} introduces four mutation
strategies implemented in \syz. At this step, \syz\ will only perform the first
three mutation strategies to vastly perturb input seeds. Evaluation by enabling
all mutation options will be given in \S~\ref{subsec:fuzzing-generation}. Since
no ``new traces'' are generated, our observation shows that the fuzzing
procedures rapidly reach to the saturation point after around 0.5 hour for all
three seeds. Still, we fuzz each seed for 3 hours to explore their full
potentials.
\begin{table}[!htbp]
  \vspace{-10pt}
	\centering
	\scriptsize
  \caption{Kernel coverage comparison using different seeds.}
  \vspace{-10pt}
	\label{tab:coverage}
	\resizebox{0.40\linewidth}{!}{
		\begin{tabular}{l|c|c|c}
			\hline
                        & $S_{rl}$ & $S_{moon}$ & $S_{def}$ \\
			\hline
			\textbf{coverage} & 25,252 & 24,932 & 14,902 \\
			\hline
		\end{tabular}
	}
  \vspace{-10pt}
\end{table}
\T~\ref{tab:coverage} reports the basic block coverage after 3-hour fuzzing.
$S_{rl}$ outperforms its competitors by achieving a higher code coverage, while
$S_{def}$ has the worst coverage (consistent with the \ms\ paper). We interpret
the results as generally encouraging; the high quality fuzzing seed generated by
\tool\ enables a practical exploration of production Linux kernels, achieving
higher coverage. We present more comprehensive evaluation in terms of coverage
and crashes in the following section.

\begin{table}[t]
  \centering
  \scriptsize
  \caption{Crashes found in kernel 5.5-rc6 and 4.17-rc4 using different seeds. To clarify
    potential confusions, ``\#Deduped Crashes'' (\textbf{\#D}) deduplicates repeated crashes
    and each reported crash has a different call stack. ``\#Unique Crashes'' (\textbf{\#U})
    reports number of deduped crashes that are only found by this particular
    seed. \textbf{\#UU} denotes unique and unknown crashes by the time of writing.}
  \label{tab:bug1}
  \label{tab:bug2}
  \resizebox{0.4\linewidth}{!}{
    \begin{tabular}{l|c|c|c|c|c|c}
      \hline
      \multirow{2}{*}{\textbf{Seeds}} & \multicolumn{3}{|c}{5.5-rc6} & \multicolumn{3}{|c}{4.17-rc4} \\
                                      & \textbf{\#D} & \textbf{\#U} & \textbf{\#UU} & \textbf{\#D} & \textbf{\#U} & \textbf{\#UU} \\
      \hline
     $S_{rl}$      & 20  & 2 & 1 & 34  & 16 & 10 \\
     $S_{moon}$    & 19  & 2 & 0 & 25  &  5 & 4  \\
     $S_{def}$     & 20  & 4 & 2 & 20  & 2  & 1  \\
      \hline
    \end{tabular}
  }
  \vspace{-20pt}
\end{table}

\begin{figure}
  \vspace{-10pt}
  \centering
  \includegraphics[width=1.0\linewidth]{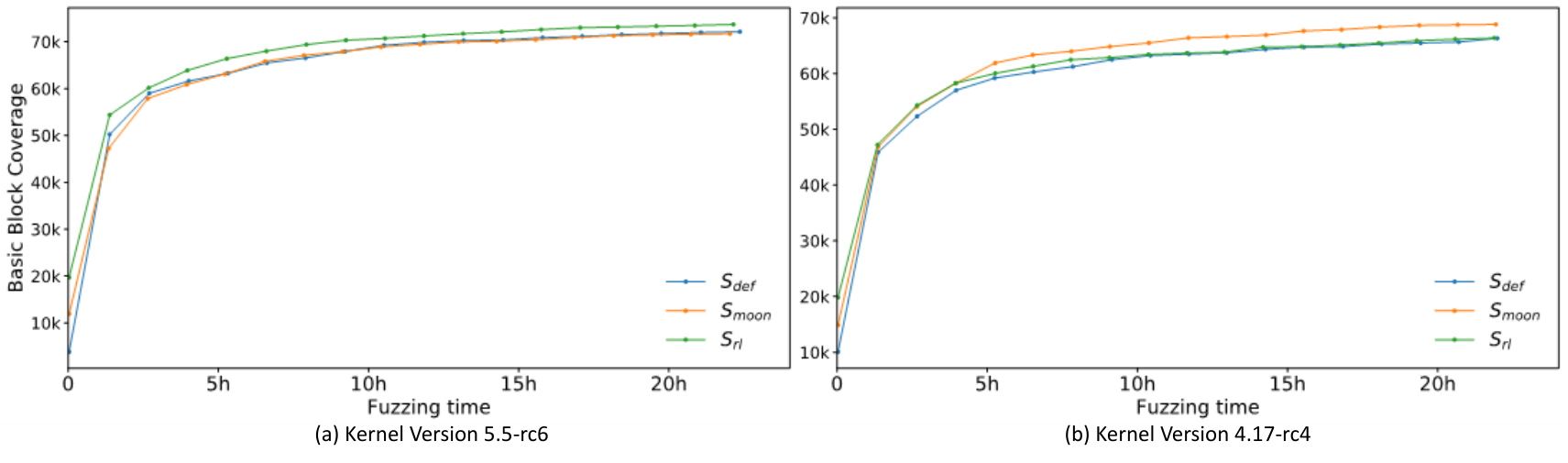}
  \vspace{-25pt}
  \caption{Code coverage with runtime trace generation enabled during 24
    hours of fuzzing.}
  \label{fig:coverage}
  \vspace{-10pt}
\end{figure}
\vspace{-5pt}
\subsubsection{Fuzzing With Runtime Trace Generation Enabled}
\label{subsec:fuzzing-generation}
As aforementioned, \syz\ can be configured to continuously generate new traces
with its thousands of hand-written rules. In this research, we enable this
feature to mimic the ``normal'' way of using \syz\ and launch a 24-hour fuzzing.
To present a comprehensive comparison, besides the latest kernel (ver. 5.5-rc6),
we also test another kernel (ver. 4.17-rc4) evaluated in the \ms\ paper.
\F~\ref{fig:coverage} reports the coverage increase during 24 hours of fuzzing.
Again, to present a fair comparison with \ms, we only generate 1,526 traces in
the evaluation. For practical usage, users can certainly generate more traces
with \tool. Within 24 hours of fuzzing, the \syz\ generates a large amount of
extra traces (close to 100K), and therefore, seeds produced by \tool\ and
\ms\ become ``insignificant'' to some extent. Nevertheless, \tool\ still
outperform its competitors, by achieving a higher coverage in kernel 5.5-rc6. In
contrast, \tool\ and \syz\ has less coverage compared to \ms\ while fuzzing the
older kernel. Note that we synthesize $S_{rl}$ w.r.t. kernel 5.5-rc6, and some
system calls in $S_{rl}$ are \textit{not} supported by kernel 4.17-rc4. Hence,
254 traces in $S_{rl}$ are directly rejected without contributing to any
coverage, presumably undermining \tool\ for kernel 4.17-rc4.

\T~\ref{tab:bug1} reports the triggered kernel crashes during the 24-hour
campaign in kernel 5.5-rc6 and 4.17-rc4. We count crashes by analyzing the crash report
provided by \syz\ and deduplicate if two crashes have identical call stacks.
While these seeds find close number of crashes from kernel 5.5-rc6, cross
comparison shows that each tool can find unique crashes that are \textit{not}
detected by others. Those cases are more interesting and are reported in the
column \textbf{\#UU} of \T~\ref{tab:bug1}. We further check whether those unique crashes
have been reported before. We carefully searched the crash stack trace from the
Linux Kernel Mailing List~\cite{lkmlurl}, Red Hat Bugzilla~\cite{redhaturl},
Lore Kernel~\cite{lorekernelurl}, and Google. We find
three (1+2) crashes in total that cannot be found anywhere, and
presumably deem \textit{unknown bugs} in kernel 5.5-rc6.
Regarding the old kernel (4.17-rc4) evaluated by \ms, \tool\ finds considerable
more crashes compared to the other seeds, and by checking each unique crash, ten
crashes exposed by \tool\ are not disclosed publicly to our best knowledge.

\section{Discussion and Future Direction to Industrial Scenarios}
\label{sec:discussion}

We believe this paper has revealed high potential of \tool\ in promoting OS
kernel fuzzing and reliability. As a starting point for future research and
engineering efforts, we list future directions of our work from the following
aspects.

\noindent \textbf{Combining with Offline Analysis.}~\tool\ trains the model from
scratch to highlight the key contribution and novelty --- synthesizing diverse
system call traces from scratch with RL. To this end, \tool\ takes code coverage
to form learning reward. Nevertheless, our evaluation in
\S~\ref{subsubsec:dependency} indicates the need of integrating dependencies
into the reward. Overall, to enhance the fuzzing performance, we plan to combine
\tool\ with offline analysis. For instance, as an offline phase, we can leverage
active learning techniques to gradually form dependencies among system calls,
and then integrate the analysis results to enhance our online fuzzing and
learning process. Moreover, we expect to leverage recent progress in AI,
particularly large language models (LLMs), to extract the dependencies among
system calls from OS kernel source code or documents. For instance,
GPT-4~\cite{gpt4} is a recently proposed LLM that can generate high-quality
text. We envision that GPT-4 can be leveraged to generate system call traces
from scratch, and we plan to explore this direction in the future.

\noindent \textbf{Feasibility Study of Fuzzing Industrial, Embedded OS
Kernels.}~From a holistic perspective, \tool\ can be easily migrated to fuzz
other OS kernels since its technical pipeline is generic and systematic. We have
tentatively explored the feasibility of migrating \tool\ to fuzz other OS
kernels, in particular, a commercial embedded OS kernel, \textit{xxxOS}, that is
being adopted in real-world, industrial sectors.\footnote{The OS kernel name is
blinded for commercial reasons.} In short, we find that fuzzing those embedded
OS kernels is not conceptually more challenging than fuzzing Linux kernels. They
however impose new technical challenges that require further research efforts.
In particular, we list the number of APIs of the two kernels in
\T~\ref{tab:api}. We observe that the number of APIs of \textit{xxxOS} is much
smaller than that of Linux. This is reasonable, as typical embedded OS kernels
are designed to be lightweight and resource-efficient. However, the small number
of APIs makes it potentially more challenging to synthesize diverse system call
traces. For instance, we find that the number of system calls that can be
invoked by a single API is much larger than that of Linux. This indicates that
the dependencies among system calls are less comprehensive, and presumably more
subtle. To combat this challenge, we plan to leverage recent progress in AI,
particularly large language models (LLMs), to extract the dependencies among
system calls from OS kernel source code or documents. We also anticipate to use
other static analysis or learning techniques, whose rational has been presented
above.

\begin{table}
  \vspace{-10pt}
  \centering
  \scriptsize
  \caption{Comparison of APIs in the Linux version evaluated in this research
   and \textit{xxxOS}.}
  \label{tab:api}
  \resizebox{0.40\linewidth}{!}{
    \begin{tabular}{l|c|c}
      \hline
                        & \textbf{Linux} & \textbf{xxxOS} \\
      \hline
     The number of APIs & 331 & 112 \\
      \hline                                                                                                                         
    \end{tabular}                                                                                                                    
  }                                                                                                                                  
\end{table}                                                                                                                          

Another major challenge is that \textit{xxxOS} is an embedded OS kernel running
on a specific hardware platform. To fuzz \textit{xxxOS}, our tentative
exploration shows that the underlying fuzzing framework, \syz, cannot be
directly used. This is reasonable, as \syz\ is designed to fuzz general-purpose
OS kernels and it largely relies on the full-system virtualization environment
to monitor the kernel execution. However, \textit{xxxOS} is an embedded OS
kernel, and it is not designed to run in a full-system virtualization
environment. Note that this could be a general and pervasive challenge when
benchmarking industrial, commercial OS kernels. To address this challenge, we
anticipate to investigate a high volume of engineering efforts to re-develop a
proper fuzzing framework for \textit{xxxOS}. We leave it as our future work.

\noindent \textbf{Securing Industrial, Embedded OS Kernels.}~From a more general
perspective, we believe that securing embedded OS kernels requires more than
fuzzing. In short, software fuzz testing mainly focuses on more obvious security
properties like memory safety, and during testing, its ``testing oracle'' is
mainly derived from system crash, hang, or other obvious symptoms. However, it
has been reported that embedded OS kernels are vulnerable to more subtle
security properties like information leakage, functional (driver) bugs, side
channel attacks, and so on. To this end, we believe it is of great importance to
develop a more comprehensive and systematic solution to secure embedded OS
kernels. Our effort and tentative exploration reported in this section is a
starting point, and we plan to explore this direction in the future. In
particular, we plan to leverage recent progress in AI, particularly LLMs, to
explore potential privacy leakage issues in embedded OS kernels. Note that
typical commercial embedded OS kernels may operate on various critical devices,
such as medical devices, automobiles, and so on. Therefore, it is of great
importance to secure them from the perspective of privacy leakage. The authors
have accumulated rich experience in handling and detecting privacy leakage bugs
using software testing, static analysis, and side channel analysis methods. We
plan to explore this direction in the future.

\vspace{-5pt}
\section{Related Work}
\label{sec:related}
\vspace{-5pt}
\noindent \textbf{OS Fuzzers.}~
Trinity~\cite{trinity} is another popular OS fuzzer.
We choose \syz\ since it is the de facto OS fuzzing
framework maintained by Google and \ms\ only uses this tool for fuzzing and
comparison. Note that Trinity~\cite{trinity} is generally not desired in our
research: Trinity is not a coverage-guided fuzzer, and, therefore, ``code
coverage'' cannot be obtained from Trinity for model training. Overall,
\tool\ does not rely on any particular OS fuzzer design, and all OS kernel
fuzzers (including Trinity) can potentially benefit from high-quality inputs
offered by \tool.

\noindent \textbf{Security Testing of OS Kernels.}~In addition to perform fuzz
testing toward OS kernel system call interface and expose memory related
vulnerabilities, existing research also aims to fine-tune the performance of
fuzz testing with respect to certain \textit{specific} OS kernel components and
vulnerabilities. For instance, Razzer~\cite{jeongrazzer} performs fuzz testing
to pinpoint race conditions within Linux kernels. Xu et al.~\cite{xu2019fuzzing}
launches effective fuzz testing toward file systems by re-scoping the mutation
target from large file image blobs into metadata blocks. Also, besides the
system call interfaces, recent research
works~\cite{song2019periscope,jiang2019fuzzing} also propose fuzz testing
framework to probe and detect bugs from the device-driver interactions with OS
kernels. Looking ahead, we leave it as one future work to integrate those
critical and specific testing tasks into \tool.

\section{Conclusion}
\label{sec:conclusion}

We have proposed a RL-based method to synthesize high-quality and diverse system
call traces for fuzzing OS kernels. The propose technique generates high-quality
traces without using software analysis or rule-based techniques. Our evaluation
shows that the synthesized system call traces engender high OS code coverage and
also reveal vulnerabilities overlooked by existing tools.

\section*{Acknowledgement}

The authors would like to thank the anonymous reviewers for their valuable
comments and suggestions. The authors also thank the engineers from the China
Academy of Industrial Internet for their help in this research.

\bibliographystyle{splncs04}
\bibliography{bib/ref,bib/analysis,bib/machine-learning}

\begin{thebibliography}{10}
\providecommand{\url}[1]{\texttt{#1}}
\providecommand{\urlprefix}{URL }
\providecommand{\doi}[1]{https://doi.org/#1}

\bibitem{trinity}
Trinity. \url{https://github.com/kernelslacker/trinity} (2018)

\bibitem{linuxvul1}
{The Top 10 Linux Kernel Vulnerabilities You Should Know}.
  \url{https://resources.whitesourcesoftware.com/blog-whitesource/top-10-linux-kernel-vulnerabilities}
  (2019)

\bibitem{lkmlurl}
Linux kernel mailing list. \url{https://lkml.org/} (2023)

\bibitem{gers1999learning}
Gers, F.A., Schmidhuber, J., Cummins, F.: Learning to forget: Continual
  prediction with lstm  (1999)

\bibitem{godefroid2008grammar}
Godefroid, P., Kiezun, A., Levin, M.Y.: Grammar-based whitebox fuzzing. In: ACM
  Sigplan Notices. vol.~43, pp. 206--215. ACM (2008)

\bibitem{godefroid2012sage}
Godefroid, P., Levin, M.Y., Molnar, D.: Sage: whitebox fuzzing for security
  testing. Communications of the ACM  \textbf{55}(3),  40--44 (2012)

\bibitem{godefroid2008automated}
Godefroid, P., Levin, M.Y., Molnar, D.A., et~al.: Automated whitebox fuzz
  testing. In: NDSS. vol.~8, pp. 151--166. Citeseer (2008)

\bibitem{godefroid2017learn}
Godefroid, P., Peleg, H., Singh, R.: Learn\&fuzz: Machine learning for input
  fuzzing. In: Proceedings of the 32Nd IEEE/ACM International Conference on
  Automated Software Engineering. pp. 50--59. ASE 2017 (2017)

\bibitem{syzkaller}
Google: Syzkaller. \url{https://github.com/google/syzkaller} (2018)

\bibitem{gupta2019deep}
Gupta, R., Kanade, A., Shevade, S.: Deep reinforcement learning for syntactic
  error repair in student programs. In: Proceedings of the thirty-third AAAI
  conference on Artificial Intelligence. AAAI 2019 (2019)

\bibitem{han2017imf}
Han, H., Cha, S.K.: {IMF}: Inferred model-based fuzzer. In: Proceedings of the
  2017 ACM SIGSAC Conference on Computer and Communications Security. pp.
  2345--2358. CCS '17 (2017)

\bibitem{redhaturl}
Hat, R.: Red hat bugzilla. \url{https://bugzilla.redhat.com/} (2023)

\bibitem{jeongrazzer}
Jeong, D.R., Kim, K., Shivakumar, B., Lee, B., Shin, I.: Razzer: Finding kernel
  race bugs through fuzzing. In: Razzer: Finding Kernel Race Bugs through
  Fuzzing. IEEE (2019)

\bibitem{jiang2019fuzzing}
Jiang, Z.M., Bai, J.J., Lawall, J., Hu, S.M.: Fuzzing error handling code in
  device drivers based on software fault injection (2019)

\bibitem{lorekernelurl}
Kernel, L.: Lore kernel. \url{https://lore.kernel.org/lists.html} (2023)

\bibitem{mnih2013playing}
Mnih, V., Kavukcuoglu, K., Silver, D., Graves, A., Antonoglou, I., Wierstra,
  D., Riedmiller, M.: Playing atari with deep reinforcement learning. arXiv
  preprint arXiv:1312.5602  (2013)

\bibitem{gpt4}
OpenAI: Gpt-4. \url{https://openai.com/research/gpt-4} (2023)

\bibitem{pailoor2018moonshine}
Pailoor, S., Aday, A., Jana, S.: Moonshine: Optimizing os fuzzer seed selection
  with trace distillation. In: UNISEX Security (2018)

\bibitem{alexandre2014opt}
Rebert, A., Cha, S.K., Avgerinos, T., Foote, J., Warren, D., Grieco, G.,
  Brumley, D.: Optimizing seed selection for fuzzing. In: Proceedings of the
  23rd USENIX Conference on Security Symposium. p. 861–875. SEC’14, USENIX
  Association, USA (2014)

\bibitem{si2018nips}
Si, X., Dai, H., Raghothaman, M., Naik, M., Song, L.: Learning loop invariants
  for program verification. In: Advances in Neural Information Processing
  Systems (NeurIPS) (2018)

\bibitem{silver2016mastering}
Silver, D., Huang, A., Maddison, C.J., Guez, A., Sifre, L., Van Den~Driessche,
  G., Schrittwieser, J., Antonoglou, I., Panneershelvam, V., Lanctot, M.,
  et~al.: Mastering the game of go with deep neural networks and tree search.
  nature  \textbf{529}(7587), ~484 (2016)

\bibitem{skybox2019vul}
Skybox: {2019 Vulnerability and Threat Trends}.
  \url{https://lp.skyboxsecurity.com/rs/440-MPQ-510/images/Skybox_Report_Vulnerability_and_Threat_Trends_2019.pdf}
  (2019)

\bibitem{song2019periscope}
Song, D., Hetzelt, F., Das, D., Spensky, C., Na, Y., Volckaert, S., Vigna, G.,
  Kruegel, C., Seifert, J.P., Franz, M.: {PeriScope}: An effective probing and
  fuzzing framework for the hardware-os boundary. In: NDSS (2019)

\bibitem{szepesvari2010algorithms}
Szepesv{\'a}ri, C.: Algorithms for reinforcement learning. Synthesis lectures
  on artificial intelligence and machine learning  \textbf{4}(1),  1--103
  (2010)

\bibitem{10.5555/1404500}
Takanen, A., DeMott, J., Miller, C.: Fuzzing for Software Security Testing and
  Quality Assurance. Artech House, Inc., USA, 1 edn. (2008)

\bibitem{wang2017skyfire}
Wang, J., Chen, B., Wei, L., Liu, Y.: Skyfire: Data-driven seed generation for
  fuzzing. In: 2017 IEEE Symposium on Security and Privacy (SP). pp. 579--594.
  IEEE (2017)

\bibitem{xu2019fuzzing}
Xu, W., Moon, H., Kashyap, S., Tseng, P.N., Kim, T.: Fuzzing file systems via
  two-dimensional input space exploration. In: 2019 IEEE Symposium on Security
  and Privacy (SP). pp. 818--834. IEEE (2019)

\end{thebibliography}
\end{document}